# Defect emission and its dipole orientation in layered ternary ZnIn$_2$S$_4$ semiconductor

Rui Wang [1,Δ], Quan Liu [3,Δ], Sheng Dai [5], Chao-Ming Liu [1,7], Yue Liu [1], Zhao-Yuan Sun [1], Hui Li [4], Chang-Jin Zhang [4], Han Wang [5,*], Cheng-Yan Xu [5], Wen-Zhu Shao [1,*], Alfred J. Meixner [3], Dai Zhang [3,*], Yang Li [1,2,*] and Liang Zhen [1,2,6]

[1] *School of Materials Science and Engineering, Harbin Institute of Technology, Harbin 150001, China*

[2] *MOE Key Laboratory of Micro-Systems and Micro-Structures Manufacturing, Harbin Institute of Technology, Harbin 150080, China*

[3] *Institute of Physical and Theoretical Chemistry, Eberhard Karls University Tübingen, Tübingen 72076, Germany*

[4] *School of Physical Science and Technology, Center for Transformative Science, ShanghaiTech University, Shanghai 201210, China*

[5] *Institutes of Physical Science and Information Technology, Anhui University, Hefei 230601, China*

[6] *Sauvage Laboratory for Smart Materials, School of Materials Science and Engineering, Harbin Institute of Technology (Shenzhen), Shenzhen 518055, China*

[7] *Laboratory for Space Environment and Physical Sciences, Harbin Institute of Technology, 150001 Harbin, China*

E-mails: liyang2018@hit.edu.cn (Y.L.); wzshao@hit.edu.cn (W.Z.S.); wanghan3@shanghaitech.edu.cn (H.W.); dai.zhang@uni-tuebingen.de (D.Z.)

**Abstract:** Defect engineering is promising to tailor the physical properties of two-dimensional (2D) semiconductors for function-oriented electronics and optoelectronics. Compared with the extensively studied 2D binary materials, the origin of defects and their influence on physical properties of 2D ternary semiconductors have not been clarified. In this work, we thoroughly studied the effect of defects on the electronic structure and optical properties of few-layer hexagonal ZnIn$_2$S$_4$ via versatile spectroscopic tools in combination with theoretical calculations. It has been demonstrated that the Zn-In anti-structural defects induce the formation of a series of donor and acceptor levels inside the bandgap, leading to rich recombination paths for defect emission and extrinsic absorption. Impressively, the emission of donor-acceptor pair (DAP) in ZnIn$_2$S$_4$ can be significantly tailored by electrostatic gating due to efficient tunability of Fermi level ($E_f$). Furthermore, the layer-dependent dipole orientation of defect emission in ZnIn$_2$S$_4$ was directly revealed by back focal plane (BFP) imagining, where it presents obviously in-plane dipole orientation within a dozen layers thickness of ZnIn$_2$S$_4$. These unique features of defects in ZnIn$_2$S$_4$ including extrinsic absorption, rich recombination paths, gate tunability and in-plane dipole orientation will definitely benefit to the advanced orientation-functional optoelectronic applications.

**Keywords:** ZnIn$_2$S$_4$, defect, photoluminescence, back focal plane imaging, dipole orientation

**Introduction**

Atomically thin character of two-dimensional (2D) semiconductors makes them sensitive to intrinsic and extrinsic perturbations from thermal equilibrium, growth and processing, leading to the appearance of structural defects, including the vacancy, anti-site, edges and grain boundary.[1-3] These defects have great ability to strongly tailor the physical and chemical properties of 2D materials, which in turn enriches the functionalities of nano-devices. During the past years, many efforts[4] have been taken to reveal the origin of defects in a representative of 2D semiconductors, transition metal dichalcogenides (TMDs), and the influence of defects on the electronic structure has been clearly revealed by microscopic[5], spectroscopic[3, 6, 7], transport[8-10] and ab initio techniques[2, 11]. Defect engineering of ternary compounds is considered more promising than the widely studied binary TMDs, because ternary 2D systems have multiple degrees of freedom to tune their physical properties through stoichiometric variation and exhibits additional unique properties that binary 2D systems do not possess. Therefore, the investigation of ternary 2D systems via defect engineering is of great significance.

As a rediscovered layered semiconductor, ternary chalcogenide $ZnIn_2S_4$ has attracted broad interest due to their unique electronic structure and physical properties.[12-14] On one hand, the presence of rich polymorphs in $ZnIn_2S_4$ endows it with a diverse range of fascinating performance. On the other hand, the direct bandgap with wide tunability in the visible spectrum and appropriate locations of conduction/valance bands make it promising for light-harvesting device applications, such as photodetection[15], solar cells[16], light-emitting diodes[17] and photocatalysis[18]. Similar to the case in ternary TMDs, defects have great influence on the physical properties of atomically thin character of 2D ZnIn$_2$S$_4$. For instance, it has been theoretically predicted that the introduction of S[19] and Zn[20] vacancies leads to the formation of defect levels below the conduction band of ZnIn$_2$S$_4$. Experimentally, these vacancies can efficiently accelerate the separation of photo-generated electron-hole pairs, prolong the lifetime of carriers and dramatically improve the photocatalysis performance in defective ZnIn$_2$S$_4$ nanosheets and hybrid structures prepared by chemical approaches. Although the significance of defects in ZnIn$_2$S$_4$ has been accepted, the correlation between the origin of defects and physical properties of 2D ZnIn$_2$S$_4$ has not been clearly explored.

In this work, for the first time, we thoroughly studied the origin of defects in exfoliated monolayer and few-layer hexagonal ZnIn$_2$S$_4$, and their effects on the electronic structure and optical properties via versatile spectroscopic tools in combination with theoretical calculations. ZnIn$_2$S$_4$ exhibited the layer-dependent electronic bandgap and extrinsic absorption characteristics, the energy of which decrease by increasing the thickness. Moreover, the rich donor and acceptor levels inside the bandgap and the resultant complicate recombination paths of ZnIn$_2$S$_4$, such as band-to-band transition, donor-acceptor pair (DAP) recombination, acceptor-conduction band recombination were carefully distinguished by combining absorption, power and temperature dependent photoluminescence (PL), deep level transient spectroscopy (DLTS) with theoretical calculations. Besides, it is discovered that the emission of DAP in ZnIn$_2$S$_4$ under the below-bandgap excitation can be significantly tailored by electrostatic gating due to the *n*-type semiconducting characteristics and efficient tunability Fermi level ($E_\text{f}$) of ZnIn$_2$S$_4$. Finally, the layer-dependent dipole orientation of defect emission was directly revealed by back focal plane (BFP) imagining, which show obviously in-plane dipole orientation within a dozen layers thickness of ZnIn$_2$S$_4$.

**Results and Discussion**

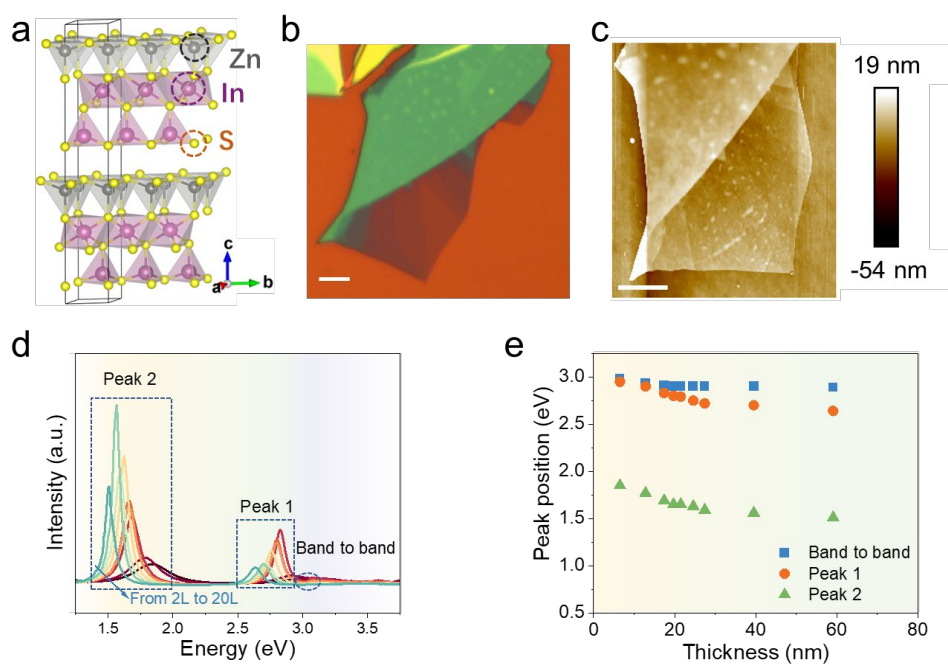

Figure 1. Electronic structure and layer-dependent absorption of ZnIn$_2$S$_4$ flakes. (a) Crystalline structure of hexagonal ZnIn$_2$S$_4$. (b) Optical image and (c) AFM topography of a flake with different thicknesses. Scale bars in (b) and (c) is 5 μm. (d) Layer-dependent absorption of the flakes with different thicknesses. (e) Extracted layer-dependent intrinsic and extrinsic absorption peaks of ZnIn$_2$S$_4$.

**Figure 1a** shows the crystalline structure of hexagonal ZnIn$_2$S$_4$, which belongs to

*P63mc* space group with lattice constants of *a*=3.85 Å, *b*=3.85 Å and *c*=24.68 Å, respectively, the electronic structure of which was further verified by X-ray diffraction (XRD) and transmission electron microscopy (TEM) measurements (**Figures S1 and S2**). **Figure 1b** and **c** shows the optical image and AFM topography of a representative exfoliated flake with different thicknesses. The thickness of a monolayer $ZnIn_2S_4$ is determined to be 2.8 nm (**Figure S3**), in agreement with the theoretical value of ~ 2.5 nm. **Figure 1d** presents the layer dependent absorption spectra of $ZnIn_2S_4$ ranging from 2 L to ~ 20 L. The peak located at ~ 3 eV results from the intrinsic absorption while another two broad peaks (2.5 ~ 3 eV and 1.5 ~ 1.9 eV) correspond to the extrinsic absorption. The optical bandgap ($E_g$) of $ZnIn_2S_4$ with different thicknesses was extracted by using the Kubelka-Munk function, and the results shown in **Figures 1e** and **S4**. It is observed that the flakes exhibit the layer-dependent characteristics of electronic bandgap, where the bandgap of $ZnIn_2S_4$ decreases from 2.98 to 2.88 eV when the thickness changes from from 2 L to ~ 20 L. It is worth noting that the flakes exhibit strong extrinsic absorption, indicating that the defect levels inside the bandgap have great contributions to the photon absorption, which is quite different from the case in other 2D materials (such as $MoS_2$, $WSe_2$) where the defect levels act as recombination paths, but have no contribution to absorption.[21-23]

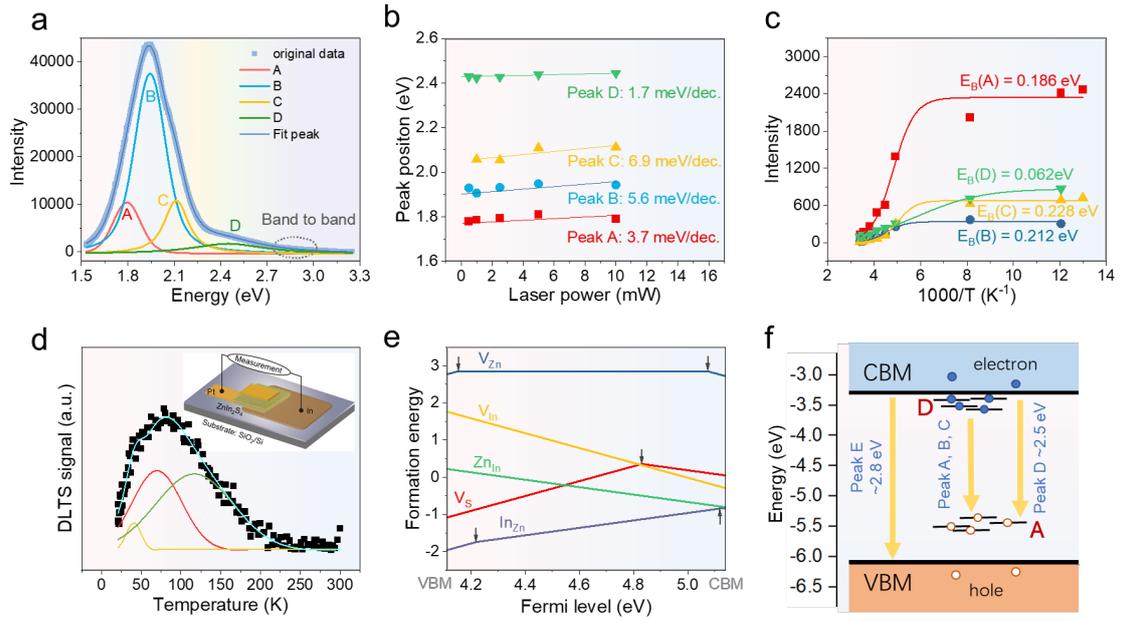

Figure 2. Photoluminescence performance by above bandgap excitation and rich recombination paths in ZnIn$_2$S$_4$ flakes. (a) Photoluminescence spectra at 300 and 77 K excited by 325 nm. (b) Integrated PL intensities of main peaks as a function of excitation power. (c) The integrated PL intensities of the four peaks (marked by A, B, C and D) as a function of temperature. (d) DLTS signal of a ZnIn$_2$S$_4$ device. Inset shows the schematics of an asymmetric ZnIn$_2$S$_4$ device for DLTS. (e) The formation energies of diverse point defect in ZnIn$_2$S$_4$. (f) Schematic diagrams of defect states inside the bandgap.

Given that luminescence can reflect the recombination behavior of defect states, we then studied the photoluminescence performance of the flakes. The flakes were excited with above-bandgap excitation via a laser wavelength of 325 nm (3.8 eV). **Figure 2a** shows the PL spectra of the flake at 300 K, where a wide peak at 1.6 ~ 2.9 eV and a shoulder (namely E) around 2.8 eV are observed. It is noted that this shoulder is more prominent at 77 K, which can be assigned to the band-to-band transition, consistent with the absorption characteristics. The broad PL peak at 300 K was fitted by Voigt function, where four peaks located at 1.78, 1.91, 2.06, 2.43 eV (namely A, B, C, D) were observed (**Figure 2a**). To reveal the origin of the corresponding recombination paths, the power-dependent PL intensity of the peaks are extracted, as presented in **Figures 2b** and **S5**, and two features are observed. One is that the power-law index of all four peaks is lower than 1, suggesting that they are from the defect emissions. The other is that the energy of the peaks (A, B, C) appear blueshifts with increasing the power intensity, which is one of the characteristics of donor-acceptor pairs (DAP) recombination luminescence.[24]

The peak energy for DAP recombination luminescence is expressed as

$$h\nu(R) = E_g - (E_D + E_A) + e^2/4\pi\varepsilon R + f(R) \qquad (1)$$

where $E_g$, $E_D$, and $E_A$ are the band gap, the donor and acceptor ionization energies, respectively, $R$ denotes the distance between the donor and acceptor defects, and $\varepsilon$ is a static dielectric constant. Here, $-e^2/4\pi\varepsilon R$ represents the Coulomb interaction energy between the donor and acceptor. In addition, $f(R) = \dfrac{-6.5e^2}{\varepsilon_0 R}(\dfrac{a_B}{R})^5$ represents the Van der Waals polarization between donor and acceptor, where $a_B$ is the Bohr radius between donor and acceptor. With the increase of the excitation power, the resultant high density of excited charge carriers leads to a stronger Coulomb interaction between DAPs and a decrease in the distance between these defects. Consequently, DAP recombination luminescence energy shifts to higher energy when the excitation power density increases, as displayed in **Figure 2b**.

We further conducted the temperature-dependent PL measurements, and the results are shown in **Figures 2c** and **S6**. It is observed that five peaks are clearly distinguished at low temperature. Given that the peak located at around 2.8 eV is assigned to be band to band transition, we focus on the broad peak at 1.5 ~ 2.5 eV. The integrated intensities of the four peaks as a function of temperature are shown in **Figure 2c**, and can be fitted by an Arrhenius equation[25]:

$$I(T) = \frac{I_0}{1 + A\exp(-E_B/k_B T)} \quad (2)$$

where $I_0$, $E_b$, $k_B$ are the PL intensity at 0 K, the thermal activation energy, and the Boltzmann constant, respectively. The binding energy of the four peaks (peaks A to D) are determined to be 0.186, 0.212, 0.228 and 0.062 eV, respectively, large than the thermal energy (26 meV) at room temperature, further suggesting the strong charge trapping ability of defect levels. Given that $ZnIn_2S_4$ is n-type semiconductor, the acceptor level may be deeper than the donor level, and hence these obtained activation energies (peaks A to C) are considered as donor levels. Therefore, the donor levels extracted by photoluminescence behavior are determined to be 0.186, 0.212, and 0.228 eV below the conduction band minimum, respectively.

Apart from temperature-dependent PL measurements, Deep Level Transient Spectroscopy (DLTS) also serves as a valuable tool to reveal the defect levels in semiconductors.[26] A Schottky diode was fabricated, as depicted in the inset of **Figure 2d**, and the electrical performance are displayed in **Figure S7**. As shown in **Figure 2d**, the DLTS signal of $ZnIn_2S_4$ has a broadened peak, which represents the majority carrier (electron) trap in the material due to the n-type semiconducting characteristics of $ZnIn_2S_4$. By splitting the DLTS signal via Gaussian fitting, the temperature values of the three peaks are 40, 69 and 117 K respectively, and the energy level positions of these three signal peaks are estimated to be located at 0.076, 0.138 and 0.243 eV below the conduction band, respectively. It is worth noting that the energy values of defect levels extracted from DLTS measurements is consistent with the values from temperature-dependent PL measurements, suggesting that at least three defect levels exist below the conduction band of $ZnIn_2S_4$. According to the bandgap of $ZnIn_2S_4$, DAP luminescence energy and the extracted donor levels, the acceptors levels are estimated to be 0.83, 0.68, 0.52 eV above the valence band maximum of $ZnIn_2S_4$.

Next, we intend to explain the origin of defect levels (donors and acceptors) in $ZnIn_2S_4$. It has been reported previously that bulk $ZnIn_2S_4$ with different polymorphs has rich intrinsic defects or disorders, especially vacancies and anti-structural defects (In-Zn exchange), leading to rich electron trapping levels and acceptors inside the bandgap of $ZnIn_2S_4$. To further investigate whether these defects have contributions to the formation of energy levels inside the bandgap, we theoretically calculated the formation energies of vacancy ($V_{Zn}$, $V_S$, $V_{In}$) and anti-structural defects ($Zn_{In}$, $In_{Zn}$) by first-principles calculations with the accurate hybrid functions. **Figure 2e** shows the dependence of defect formation energies on chemical potential. It is worth noting that the calculated bandgap of $ZnIn_2S_4$ is estimated to be 1.12 eV, lower than the experimental value. It is found that Zn vacancy ($V_{Zn}$) is an amphoteric defect which can act as both acceptor with defect level at 0.042 eV above the VBM and donor at 0.068 eV below the CBM. However, the formation energy of $V_{Zn}$ is largest, and hence Zn vacancy is not easy to exist. Regarding $V_{In}$ defect, it has relatively low formation energy without any defect-level in the bandgap. Compared with $V_{Zn}$ and $V_{In}$, sulfur vacancy ($V_S$) acts as a deep donor with defect level at 0.31 eV below the CBM, while the formation energy of $V_S$ is quite low under n-type condition, indicating that $V_S$ is more likely to exist. Similarly, it is found that $Zn_{In}$ creates no defect level inside the bandgap, whereas $In_{Zn}$ is an amphoteric defect with an acceptor level (acceptor) at 0.11 eV above the VBM and a donor level at 0.02 eV below the CBM.

Experimentally, the anti-structural defects and S vacancies are observed from high-angle annular dark-field (HAADF) scanning trans-mission electron microscopy (STEM) and energy dispersive X-ray spectrometer (EDX) measurements, respectively, as shown in **Figures S8** and **S9**. According to the theoretical calculations above mentioned, these two kinds of defects can lead to the formation of energy levels inside the bandgap of $ZnIn_2S_4$, which is in agreement with previous reports about bulk counterparts[27-29], where the intrinsic defects in bulk $ZnIn_2S_4$ induce exponentially distributed shallow donor states and acceptors. Therefore, in the following, we will explain the rich recombination paths, as depicted in **Figure 2f**. The D and A level indicate the donor and the acceptor levels originating from $In_{Zn}$ defect respectively, while the deep level marked by S is attributed from the S vacancy. After above-bandgap excitation, there are several kinds of radiative recombination paths, marked by A to E (**Figure S6**). Peak E located at 2.75 eV is attributed to the band-to-band recombination, and peaks A-C is located at 1.6 ~ 2.5 eV for the DAP recombination (free electrons are captured by trap levels (D levels), and subsequently make radiative transitions to the A levels). Peak D at 2.5 eV is associated with the acceptor state involved in the donor-acceptor luminescence transition that induces electron transition from the A levels to the CBM.

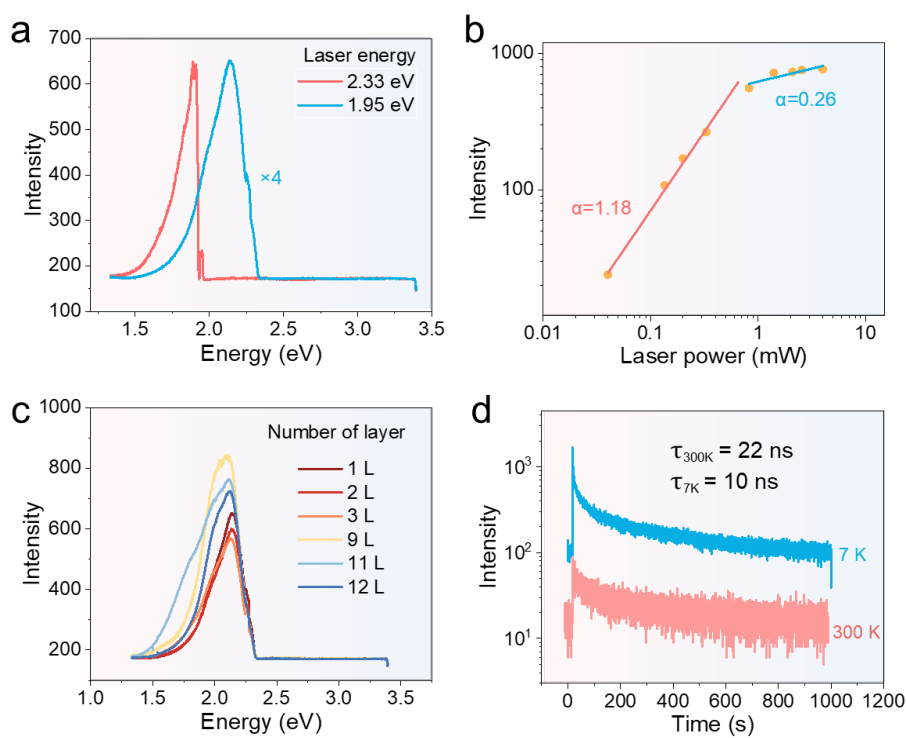

Figure 3. Photoluminescence performance of ZnIn$_2$S$_4$ flakes with below-bandgap excitation. (a) Wavelength-dependent PL spectra ZnIn$_2$S$_4$. (b) Power-dependent PL spectra of ZnIn$_2$S$_4$ excited by laser energy of 2.33 eV. (c) Layer-dependent PL spectra and (d) Time-resolved PL of the flake at 7 and 300 K excited by ps laser with 2.33 eV.

As mentioned that the flake exhibits strong extrinsic absorption and defect emission, we then excited the flakes with below-bandgap excitation in order to further understand the origin of defects. **Figure 3a** shows the PL spectra of the flake with a thickness of 3-layer excited by a laser with wavelengths of 2.33 and 1.95 eV. It is found that the flake exhibits a wide peak under excitation by 2.33 and 1.95 eV, while the PL intensity excited by 1.95 eV is 4 times stronger than that excited by 2.33 eV. From the power-dependent PL spectra shown in **Figures S10** and **3b**, the power-law behavior ($I = P^\alpha$) changes from superlinear ($\alpha = 1.18$) to sublinear ($\alpha = 0.26$) variation, and the PL intensity eventual saturates at large excitation power intensity. Given the feature of DAP recombination, where photoexcitation produces a non-ionized acceptor and possibly a non-ionized donor, the maximum value of α should be 2.[27] In our case, the value of α is less than 2, because not every photoexcited electron can be trapped by a specific donor-acceptor pair under excitation of 2.33 eV. At high excitation power intensity, due to the trap saturation, the carrier trapping and resultant DAP recombination are no longer effective for most free electrons, leading to the saturation of PL intensity. Therefore, under 2.33 eV excitation, the electrons can be excited from A level to the CBM or donor levels, and then captured by the traps, leading to the efficient DAP recombination. In contrast, the much lower PL intensity of $ZnIn_2S_4$ excited by 1.95 eV is due to inefficient excitation and trapping of carriers.

Whether the DAP recombination depends on the layer number of $ZnIn_2S_4$? **Figure 3c** shows the thickness-dependent PL of the flakes excited by 2.33 eV. With increasing the thickness from 2.8 to 36 nm, the PL peak energy exhibits a slightly redshift, suggesting that the energy levels of defects change slightly by the thickness. However, the intensities between different thickness from 2.8 to 36 nm without big change is shown in **Figure S11**. From the time-resolved PL (TRPL) results shown in **Figure 3d**, the lifetime of the peak located at 2.1 eV is estimated to be 22 and 10 ns at room temperature and 7 K. This ultra-long lifetime also confirms the characteristics of the defect emission.

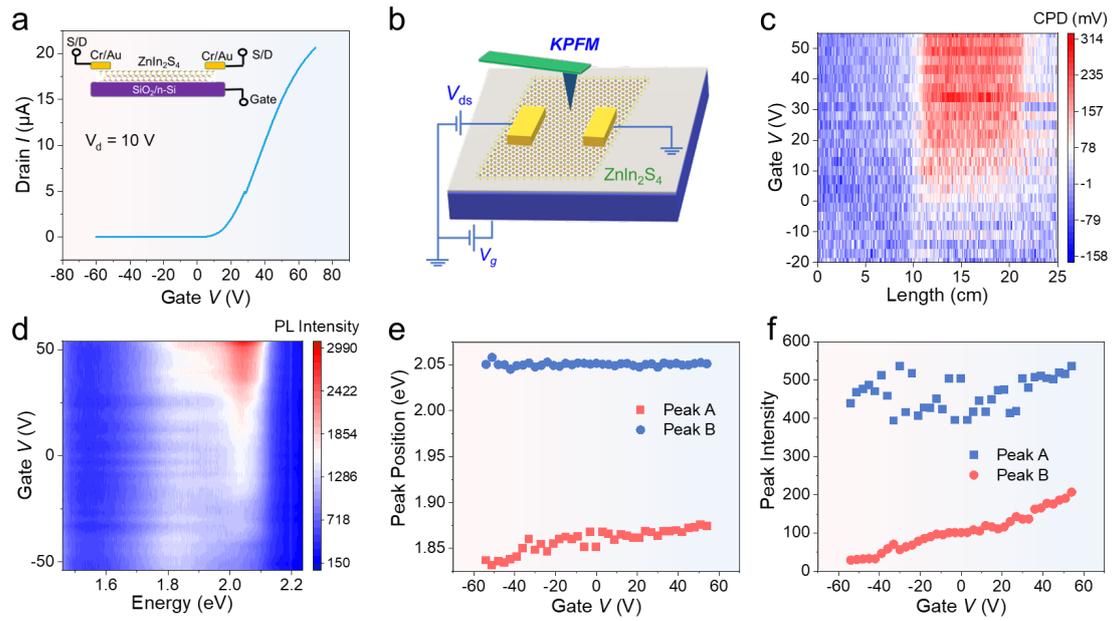

Figure 4. Gate-dependent electrical and optical performance of 8-layer ZnIn$_2$S$_4$ field effect transistor fabricated on SiO$_2$/Si substrate with Cr/Au electrode at room temperature. (a) FET performance of ZnIn$_2$S$_4$ device at room temperature with the thickness of 9-layer. Inset in (a) shows the schematic of the device. (b) Schematics of gate-dependent KPFM measurements. (c) Potential profiles along the channel at different gate voltage. (d) Gate-dependent PL intensity map at the power of 1.3 mW at 300 K with 2.33 laser excitation. (e) Peak positions and (f) peak intensities as a function of gate voltage.

Given the semiconducting characteristics of few-layer ZnIn$_2$S$_4$, we then studied its gate-dependent electrical and optical properties. **Figure 4a** shows the transfer curve of the device at a source-drain voltage of 10 V at room temperature, from which we found that the device presents electron-conduction characteristics with ON/OFF ratio up to $10^7$.

To further explore the gate-tunable $E_f$ of ZnIn$_2$S$_4$, we carried out gate-dependent Kelvin probe force microscopies (KPFM) measurements. **Figure 4b** shows the schematics of KPFM measurements, where both source and drain electrodes were grounded and $V_g$ was applied on the heavily n-doped Si substrate. Surface potential profiles measured along Au-ZnIn$_2$S$_4$-Au under different $V_g$ are shown in **Figure 4c**. Due to the difference in the work function between Au and ZnIn$_2$S$_4$, a gradual change in contact potential difference (CPD) is observed at Au-ZnIn$_2$S$_4$ junction. The surface potential of Au-ZnIn$_2$S$_4$-Au varies by gate voltages due to the presence of the unscreened long-range electrostatic interaction between the tip and $V_g$.[30, 31] It is worth noting that the surface potential of metals does not change with $V_g$, therefore, the relative difference in the CPD of Au-ZnIn$_2$S$_4$ ($\Delta V_{CPD}$) can reflect the effect of $V_g$ on the surface potential of ZnIn$_2$S$_4$. As can be seen in **Figure 4c**, $\Delta V_{CPD}$ increases significantly from 0 to 230 mV when $V_g$ varies from -20 to 55 V, indicating that the $E_f$ of the flake can be strongly tuned by electrostatic gating.

Considering the large $E_f$ shift by electrostatic gating, we next investigated whether back gate voltage can control the spectral emission properties of ZnIn$_2$S$_4$. **Figure 4d** shows in a contour plot the photoluminescence intensity as a function of $V_g$ with 2.33 eV excitation. The fitted peaks were selectively displayed in **Figure S12**. We clearly notice that the intensity increases monotonically as $V_g$ decreases. The peak positions at different gate voltages are depicted in **Figure 4e**, where the peak energy exhibits a slight redshift with increasing the gate voltage, suggesting that the energy levels of defects are not sensitive to the electrical field. From Figure 4f, the peak intensities increase significantly with $V_g$ increasing. It is evident that the electrical environment determines the activity of defect emission, duing to defect stability modified by changes to the local potential energy landscape induced by the applied voltage.[32] It is well known that large donor and acceptor doping concentrations can lead to the intense DAP recombination.[24, 33] In n-type ZnIn$_2$S$_4$ flake, the high electron density at donor states can be achieved due to the high energy level of $E_f$.[34] With increasing the gate voltages, the flake is more *n*-doped and generates high density at donor states, leading to the enhanced PL emission.

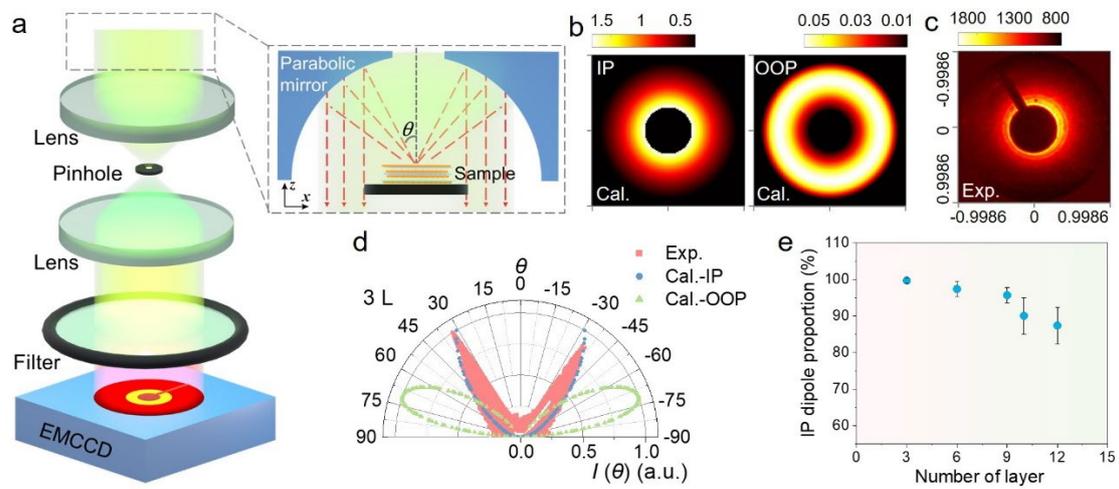

Figure 5. Thickness-dependent defect emission of the flakes via back focal plane imaging. (a) Schematics of BFP measurements. (b) Calculated and (c) experimental IP and OOP k-space emission patterns of the flakes with 3-layer thickness. (d) Measured and calculated far-field emission patterns as a function of the emission angle $\theta$ for the flakes in (b). (e) Layer-dependent in-plane dipole orientation proportion of ZnIn$_2$S$_4$ flakes.

Given that ZnIn$_2$S$_4$ has great potential for optoelectronic applications via defect engineering, it is of great importance to reveal the dipole orientation of defect emission. Back focal plane (BFP) and techniques help to measure wavevector distribution[35], where every point in the image plane corresponds to a distinct angle ($\theta$) of emission or the photo momentum along the certain $\theta$ ($k = k_0 \cdot n \cdot \sin(\theta)$), where $n$ is the index of refraction and $k_0$ is the wave vector in air. **Figure 5a** shows the schematics of BFP measurements via parabolic mirror setup. The numerical aperture (NA) is 0.986, and more experimental details is described in method of supplementary information. An analytical model put forward by Andreas Lieb et al.[36] to simulate ***k***-space emission pattern of $ZnIn_2S_4$ to confirm the dipole orientation in defect recombination. In order to test the accuracy of our measurement and analytical model, the dipole orientation of monolayer $WSe_2$ was measured shown in **Figure S13**, which is in consistent with its in-plane dipole characteristics.[37, 38]

**Figure 5b** displays the calculated pure IP and OOP ***k***-space emission patterns of the 3-layer flake, respectively, excited by 2.33 eV laser energy. From the emission patterns, it can be obviously observed that the intensity of OOP is about 30 times lower than IP pattern integrally, ascribing to the larger emission outcoupling of IP dipole.[39] Another feature is that the high intensity region scatters and forms a ring in OOP pattern while the IP pattern concentrates in the center albeit that there is a block (caused by sample holder) in the middle of the pattern, on account of the different field intensity distribution between the IP and OOP orientation.[38, 40] **Figure 5c** shows the measured emission orientation in 3-layer $ZnIn_2S_4$ that could be appropriately recognized as IP dipole. **Figure 5d** draws normalized intensity along $\theta$ from the measured and calculated both IP and OOP ***k***-space emission patterns for the flakes in **Figures 5b** and **c**, where the maximum intensity of measurement and simulation curves are at $\sim\theta = 30°$ and intensity is gradually decreasing with the increase of $\theta$, further confirming the IP orientation in fewer-layer ZnIn$_2$S$_4$. Moreover, the layer-dependent emission orientation in ZnIn$_2$S$_4$ ranging from 2-layer to 14-layer was investigated, and the results are summarized in **Figures 5e** and **S14**, from which we conclude that the defect emission in ZnIn$_2$S$_4$ exhibits a mainly IP dipole component. This extrinsic absorption and IP dipole characteristics of defect states in few-layer ZnIn$_2$S$_4$ will definitely benefit to the advanced orientation-functional optoelectronic applications.

**Conclusions**

In summary, we have thoroughly investigated the origin of defects in mechanically exfoliated few-layer hexagonal ZnIn$_2$S$_4$, and their effects on the electronic structure and optical properties. ZnIn$_2$S$_4$ exhibited the layer-dependent bandgap and extrinsic absorption characteristics. Secondly, we discussed the abundant donor and acceptor levels of ZnIn$_2$S$_4$ inside the bandgap and carefully distinguished the resultant complicate recombination paths of band-to-band transition, donor-acceptor pair (DAP) recombination and acceptor-conduction band recombination. Besides, it is revealed that the $E_f$ of *n*-type ZnIn$_2$S$_4$ semiconductor has efficient gate tunability, and the emission of DAP in ZnIn$_2$S$_4$ can be significantly controlled by static electric gating even under laser excitation below the bandgap. Finally, through the back focal plane (BFP) imagining, it directly revealed that layer-dependent dipole orientation of defect emission, determining the almost pure IP orientation within a dozen layers thickness of ZnIn$_2$S$_4$ in aspect of defect emission. Our work demonstrates the great ability to tune the unique physical properties of 2D ternary semiconductors via defect engineering for multifunctional optical and optoelectronic applications in the future.

# Supplementary Information for

# **Defect emission and its dipole orientation in layered ternary ZnIn$_2$S$_4$ semiconductor**

Rui Wang[1,Δ], Quan Liu[3,Δ], Sheng Dai[5], Chao-Ming Liu[1,7], Yue Liu[1], Zhao-Yuan Sun[1], Hui Li[4], Chang-Jin Zhang[4], Han Wang[5,*], Cheng-Yan Xu[5], Wen-Zhu Shao[1,*], Alfred J. Meixner[3], Dai Zhang[3,*], Yang Li[1,2,*] and Liang Zhen[1,2,6]

[1] School of Materials Science and Engineering, Harbin Institute of Technology, Harbin 150001, China

[2] MOE Key Laboratory of Micro-Systems and Micro-Structures Manufacturing, Harbin Institute of Technology, Harbin 150080, China

[3] Institute of Physical and Theoretical Chemistry, Eberhard Karls University Tübingen, Tübingen 72076, Germany

[4] School of Physical Science and Technology, Center for Transformative Science, ShanghaiTech University, Shanghai 201210, China

[5] Institutes of Physical Science and Information Technology, Anhui University, Hefei 230601, China

[6] Sauvage Laboratory for Smart Materials, School of Materials Science and Engineering, Harbin Institute of Technology (Shenzhen), Shenzhen 518055, China

[7] Laboratory for Space Environment and Physical Sciences, Harbin Institute of Technology, 150001 Harbin, China

E-mails: liyang2018@hit.edu.cn (Y.L.); wzshao@hit.edu.cn (W.Z.S.); wanghan3@shanghaitech.edu.cn (H.W.); dai.zhang@uni-tuebingen.de (D.Z.)

# Methods

***Sample preparation.*** ZnIn$_2$S$_4$ flakes were obtained by mechanical exfoliation from single crystals. The Bulk ZnIn$_2$S$_4$ was grown by a chemical vapor transport method (CVT)[1]. The few-layer samples were mechanical exfoliated on SiO$_2$/Si substrate with 280 nm thickness SiO$_2$. The few-layers were judged by optical microscope (*Imager A2m*, *Zeiss*), and the number of layers were determined by AFM (*Dimension Icon-PT*, *Bruker*).

***Structural and morphology characterization***. The structure was carried out by powder X-ray diffraction (*Empyrean*) with Cu Kα radiation. The TEM measurements were conducted in *talos f200x* with 200 KV acceleration voltage. The HAADF-STEM was performed by spherical aberration corrected transmission electron microscope (AC-TEM, *FEI Themis Z*) with 80 KV acceleration voltage. The samples for TEM and AC-TEM were on the microgrid copper mesh using wetting-enabled-transfer.

The morphology characterization was proceeded in AFM through *Scanasyst* mode with *Scanasyst air* tip, where the samples were mechanical exfoliated on SiO$_2$/Si substrate with 280 nm thickness SiO$_2$. Morphology and element distribution was analysed by scanning electron microscopy (SEM, *SUPRA55*) equipped with energy dispersive X-ray detector (EDX, *Oxford*), where the sample was prepared on conductive tape.

***Characterization of optical performance.*** The microscopic absorption spectra and optical images were collected in a home-designed spectroscopy system (Gora-UVN-FL, built by Ideaoptics, Shanghai, China), with a Xeon light source between 300 nm and 1100 nm. The power-dependent and temperature-dependent PL excited by a laser with energy of 3.82 eV was carried out by *LabRAM HR Evolution*, integrated with a liquid nitrogen cryostat (*Linkam*).

The BFP was measured by a home-built scanning confocal microscope, the luminous emitted from sample were reflected by parabolic mirror[2]. The light passed through confocal system (consists of two tube lens with focal of 100 mm and a pinhole of 100 μm) and then across the filter to remove the laser energy. Finally, the BFP image was collected at electron multiplying charge coupled device (EMCCD).

***DLTS measurement.*** The DLTS was performed using the *Phystech FT-1230* system. The DLTS deduces the defect level properties by measuring the response of the capacitance to temperature using Schottky device. The initial bias is -5 V. The capacitance transients (Δ$C$) were measured with a 1 MHz DLTS spectrometer operated with a temperature scan from 20 K to 300 K with a step of 2 K. The reverse bias voltage and the pulse voltage were -5 V and 0 V, respectively. The pulse width (Tw) during the test was 1.92 ms.

***DFT calculation.*** Calculations were performed using Vienna Ab initio Simulation Package (VASP), based on density function theory (DFT). A 4 × 4 × 1 supercell with 224 atoms was employed for all calculations. The exchange-correlation effect was described within the generalized gradient approximation in the Perdew–Burke–Ernzerhof (PBE) functional and the van der Waals (vdW) interaction was described with the DFT-D3 method. Projector-augmented wave method was used to describe the interaction between ion cores and valence electrons. Plane-waves kinetic energy cutoff was set to 400 eV for geometry optimization and the total energy convergence criterion of $10^{-6}$ eV is adopted. The optimized convergence criterion for atomic coordinates was less than 0.01 eV/Å for forces on each atom. The 1 × 1 × 1 Monkhorst–Pack k-meshes were used to sample the Brillouin zone for all structures presented in this study. For total energy calculation, Heyd-Scuseria-Ernzerhof (HSE06) hybrid functional was applied and the kinetic energy cutoff was set to 300 eV.

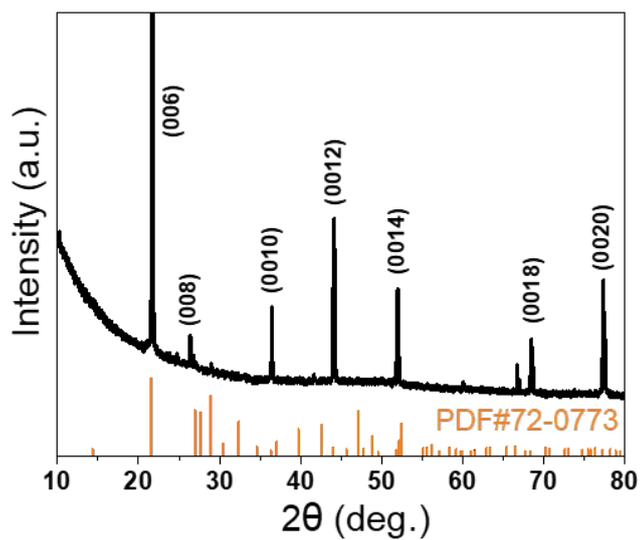

Figure S1. X-ray diffraction of bulk ZnIn$_2$S$_4$ by $\theta$-$2\theta$ scanning.

Only the (00$l$) reflections show up in the diffractogram, suggesting single [001] out-of-plane orientation. Diffraction pattern indicates that the flake is a single crystal with hexagonal structure (ICDD card *no.*72-0773)

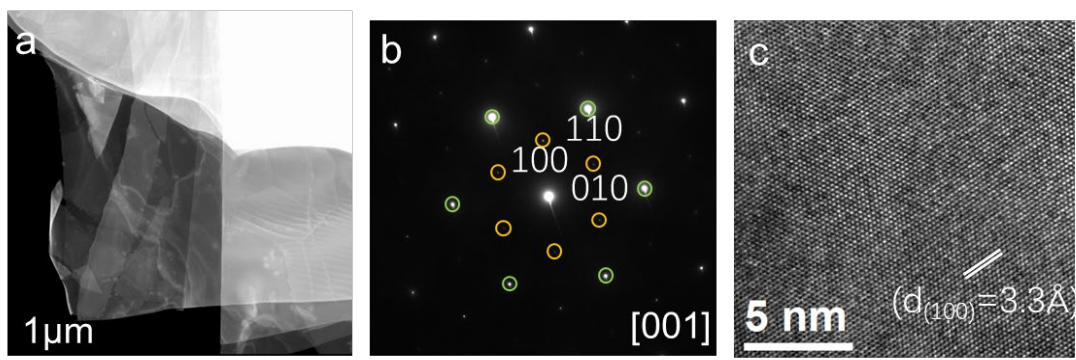

Figure S2. (a) Low-magnification TEM image, (b) the corresponding selected area electron diffraction (SAED) pattern and (c) high-resolution TEM of ZnIn$_2$S$_4$ flake.

SAED patterns distribute alternating light and dark, where dark dots are circled by yellow and light dots are circled by green. It is because that the reciprocal lattices show extinction of reflections hkl with -h+k+l≠3n, which indicates that it exhibits an ordered hexahedral structure [3]. From the high-resolution TEM image of few-layer ZnIn$_2$S$_4$, the lattice spacing of (100) is measured to be 3.3 Å.

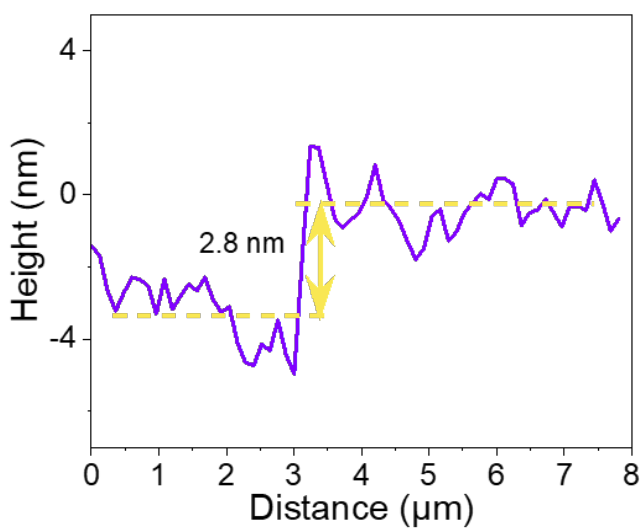

Figure S3. Heigh profile of monolayer ZnIn$_2$S$_4$ flake in Figure 1c. The single layer was measured as 2.8 nm, consistent with the theoretical value of 2.5 nm.

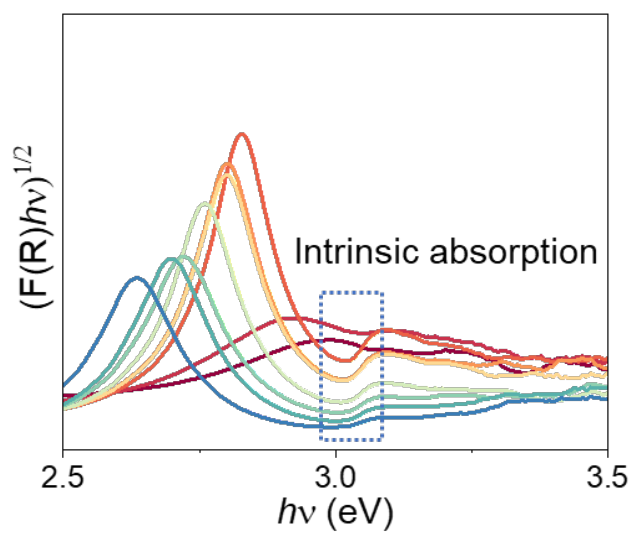

Figure S4. thickness-dependent optical absorption spectra of ZnIn$_2$S$_4$ flakes. The bandgap of ZnIn$_2$S$_4$ flakes were determined by using the Kubelka-Munk function.

We extracted the optical bandgap (E$_g$) from the optical absorption spectra by using the Kubelka-Munk function since the ZnIn$_2$S$_4$ has a direct band gap: $[F(R_\infty)h\nu]^{0.5}=A(h\nu-E_g)$. The intersection of the tangent with the x-axis can give a good approximation of the E$_g$ values of the samples, estimating to be 2.88~2.98 eV from 2 L to ~20 L

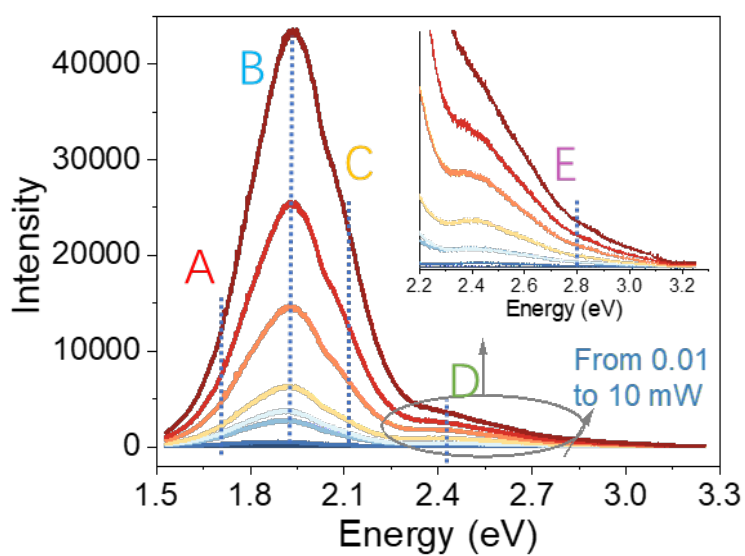

Figure S5. Power-dependent photoluminescence spectra of the ZnIn$_2$S$_4$ flake from 0.01 to 10 mV laser power with 4 extrinsic peak (A-D) and weak intrinsic peak (E).

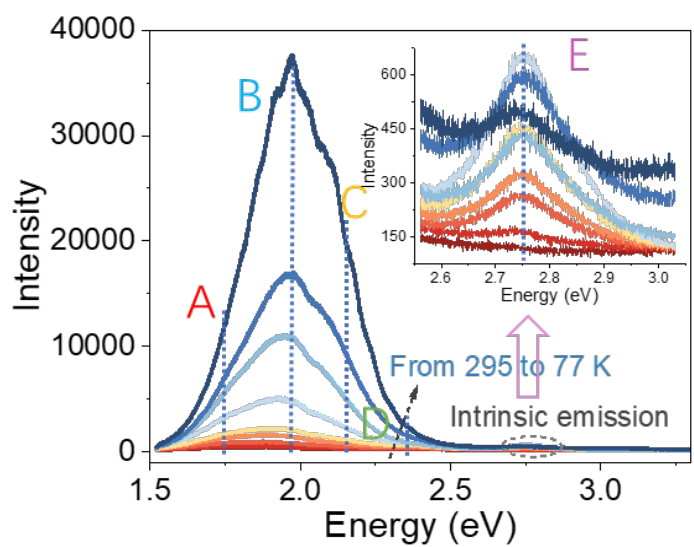

Figure S6. Temperature-dependent PL spectrum of the ZnIn₂S₄ flake, from which the intensity is increase with the decrease of temperature from 295 k to 77 k.

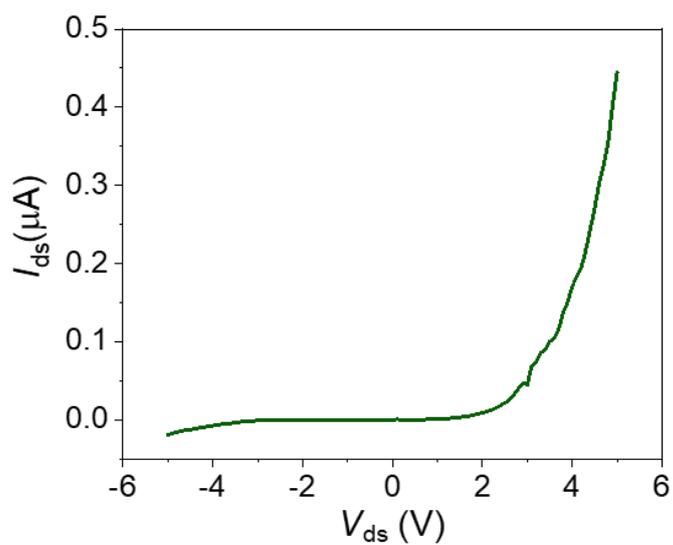

Figure S7. Electrical performance of ZnIn₂S₄ Schottky diode fabricated by indium and platinum metal electrode at room temperature.

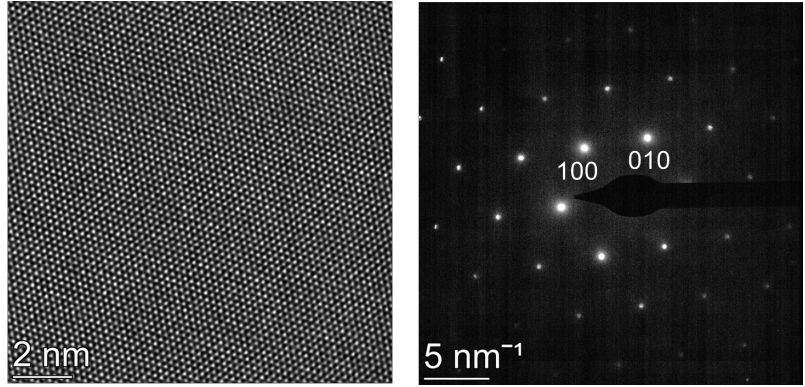

Figure S8. An AC-STEM measurements of ZnIn$_2$S$_4$. The left is quasi-disordered [4] distribution which mainly refers to the random distribution of In, Zn atom in local area, and the right is the corresponding SAED patterns.

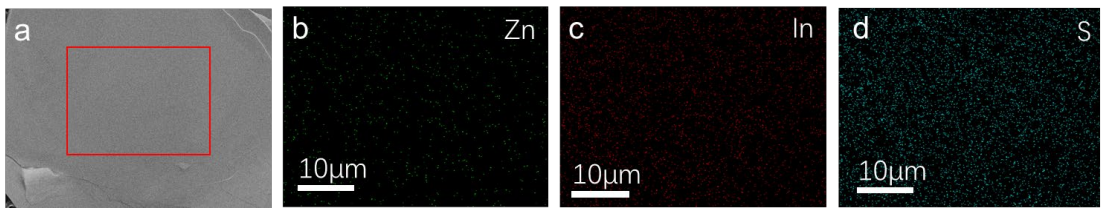

Figure S9. (a) Morphology and (b-d) EDS mapping of ZnIn$_2$S$_4$ flake under SEM with the atomic ratio of Zn, In, S, is approximately 1.22: 2: 3.78.

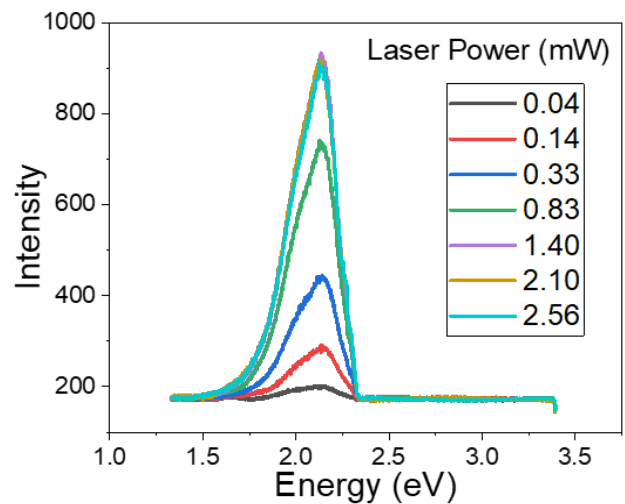

Figure S10. Power-dependent PL spectra of ZnIn$_2$S$_4$ with 3-layer under excitation of 2.33 eV laser energy at room temperature.

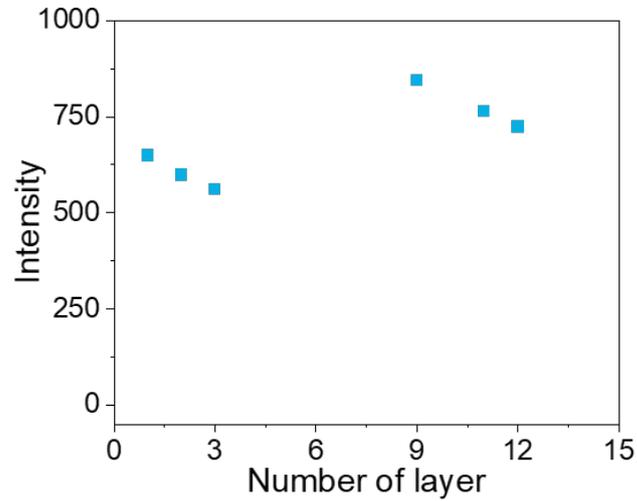

Figure S11. The PL intensity varied from 1-layer to 12-layer under excitation of 2.33 eV laser energy with the power of 0.83 mW.

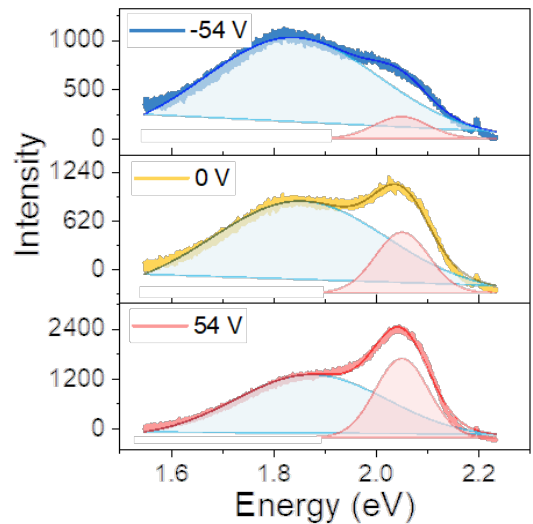

Figure S12. The selectively fitted result of PL spectra at gate voltages of -54, 0 and 54 V under the excitation of 2.33 eV laser energy. Two peaks are located at ~1.8 and ~2.1 eV, respectively.

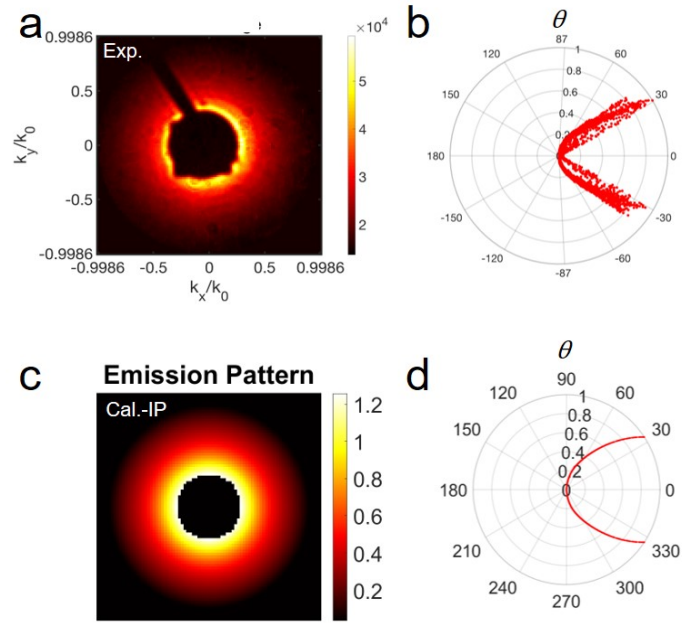

Figure S13. The measured IP orientation of ***k***-space emission pattern (a) and a function of the emission angle θ (b) of monolayer WSe$_2$. (c) and (d) are the corresponding calculated results.

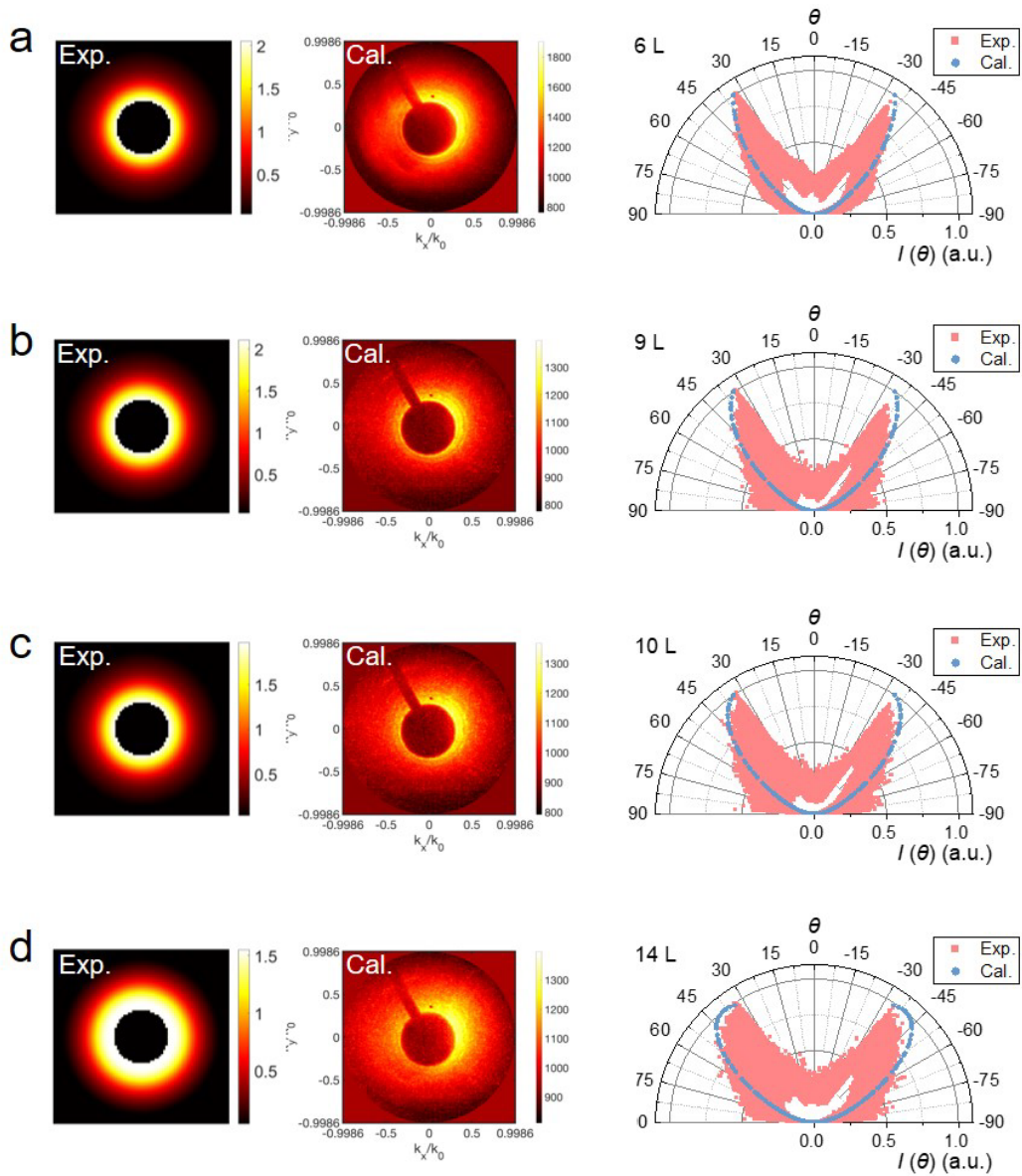

Figure S14. Thickness-dependent *k*-space emission patterns of measured and calculated images from defect emission with flakes of (a) 6-layer, (b) 9-layer, (c) 10-layer and (d) 14-layer.